\documentclass[11pt,a4paper]{article}
\usepackage{a4wide}

\begin{document}

\title{The resultant parameters of effective theory}
\author{A.~Vereshagin\thanks{Bergen University and St.-Petersburg
State University. E-mail: alexand@fi.uib.no}\ \ and
V.~Vereshagin\thanks{St.-Petersburg State University. E-mail:
vvv@av2467.spb.edu}}
\maketitle
\begin{abstract}
This is the 4-th paper in the series devoted to a systematic study of
the problem of mathematically correct formulation of the rules needed
to manage an effective field theory. Here we consider the problem of
constructing the full set of essential parameters in the case of the
most general effective scattering theory containing no massless
particles with spin
$J > 1/2$.
We perform the detailed classification of combinations of the
Hamiltonian coupling constants and select those which appear in the
expressions for renormalized
$S$-matrix
elements at a given loop order.
\end{abstract}

\section{Introduction}
\label{sec-introduction}
\mbox{}

At first glance, the concept of an effective field theory (first
formulated in
\cite{WeinEFT})
looks too general to be of practical use in computing the
characteristics of hadron scattering processes. In all the known cases
of its application (see, e.g.,
\cite{WeinEFT}, \cite{GassLeut}, \cite{Georgi}),
the authors, in fact, mostly rely on the philosophy rather than on
certain computational scheme accounting for specific features of
effective theories. The point is that such a scheme has not been
developed yet, and many questions still require answers. At the same
time, the importance of the subject is beyond question because, if
constructed, such a scheme could provide us with a tool allowing to
manage the conventionally nonrenormalizable theories (see
\cite{WeinAsySafe}).

In our previous publications
(\cite{VV} -- \cite{AVVVKS})
it was shown that, under certain conditions, it is possible to derive
quite reasonable results already from the analysis of the lowest order
amplitudes computed with a help of the most general effective
Hamiltonian constructed from local fields describing free particles
with arbitrary spins and masses. In those papers, however, many
important issues concerning the details of our approach have not been
explained. In particular, the solution to the problem of
parametrization of scattering amplitudes was declared without proof.
In this paper we discuss this issue in detail. We introduce the notion
of minimal parametrization and show that the set of minimal parameters
is quite sufficient for fixing the analytic form of arbitrary complex
graph. Moreover, it happens possible to single out those combinations
of minimal parameters which are needed to fix the form of the
amplitude of a given process at arbitrary high loop order. The latter
combinations are called as the resultant parameters. At last, we
briefly discuss the problem of ordering of infinite sums of graphs
describing the tree level amplitudes of binary processes
(this can be generalized for more involved cases) and outline a way to
construct the essential parameters -- the only ones which require
formulating the renormalization prescriptions.

\section{Preliminaries}
\label{sec-preliminaries}
\mbox{}

~First of all we need to specify the precise meaning of the term
{\em effective theory}.
It is often understood as just a theory describing physics below some
scale
$\Lambda$
(see, e.g.,
\cite{Georgi}).
In fact, this definition tacitly implies that the corresponding
perturbation series loses its meaning at energy
$E \sim \Lambda$
where a kind of new physics comes into play. We would like to stress
that we consider here just an opposite case. It is assumed that the
discussed below general effective theory does not contain any kind of
a latent inner cutoff. Owing to this, we use the term
`effective theory'
in its original meaning defined in
\cite{WeinEFT}.
Namely,
{\em we call a theory as effective if the corresponding quantum
Hamiltonian (in the interaction picture) takes a form of the
formal infinite series containing
\underline{all}
the local terms consistent with a given symmetry requirements}.

In this paper we are interested in consideration of the general
features of effective theories. Because of this reason we do not imply
the presence of any other symmetry but Lorentz invariance. The problem
of accounting for the requirements of dynamical (non-linear)
symmetries is briefly discussed below.

It is necessary to stress here that the given above definition is only
meaningful if the quantum interaction Hamiltonian can be constructed
`by hands' without any refereing to the corresponding classical
Lagrangian. This means that the canonical quantization scheme (based
on the Lagrangian) cannot be considered as the basis for constructing
the quantum effective theory because the most general form of
classical Lagrangian must contain the terms with arbitrary high powers
(and orders) of the time derivatives; in such a situation the
canonical quantization looks impracticable. Because of this reason we
rely upon the alternative -- intrinsically quantum -- scheme of
constructing an effective theory. In this scheme, developed by
S.~Weinberg in the series of papers
\cite{WeinQuant}%
\footnote{See also the Chapters 2-5 of the monograph
\cite{Weinberg}.},
the structure of the Fock space of asymptotic states is postulated,
and field operators are constructed in accordance with symmetry
properties of those states. The Hamiltonian is also postulated as the
interaction picture operator only depending on those fields and their
derivatives. The
$S$-matrix
elements are computed with the help of Dyson's formula
\begin{equation}
S_{fi} =
\langle f|
T_{\stackrel{\;}{\scriptscriptstyle\!\! W }}
exp \left\{ -i \int\limits_{}^{}H_{int} dx \right\}
|i \rangle ,
\label{1.1}
\end{equation}
where the symbol
$
T_{\stackrel{\;}{\scriptscriptstyle\!\! W }}
$
stands for Wick's T-product%
\footnote{It is explicitly covariant -- see, e.g.
\cite{Vasiliev}.}.
The noncovariant terms in the Hamiltonian and in propagators (see
\cite{Weinberg}
and the Refs. quoted therein) should be neglected -- in the case
of effective theory this does not introduce any uncertainty
because, by construction, the Hamiltonian contains
{\em all}
the terms consistent with Lorentz symmetry. This means that the
total effect of noncovariant terms might, at most, result in a
renormalization of some coupling constants.

Thus we see that Weinberg's scheme happens well suited for
constructing the effective field theory Hamiltonian. However, there is
one problem revealing itself when this scheme is used to describe the
hadron dynamics. The point is that in this scheme the Hamiltonian
contains those and only those field operators which correspond to the
states of stable particles.
{\em Weinberg's scheme is adapted to describe the scattering
processes with true stable particles solely in terms of the
corresponding creation and annihilation operators; the possibility to
describe the physics of resonances in the framework of this scheme
looks questionable.} Fortunately, this problem happens quite solvable.
The results obtained in
Ref.~\cite{Veltman} %
show that, in the case when the Hamiltonian contains the fields of
unstable particles%
\footnote{Those with masses large enought to make it possible the
decay into lighter particles.},
the formal construction
(\ref{1.1})
remains applicable. In this case it defines the
$S$-matrix
as the unitary operator on the space of stable particle states. The
fields of unstable particles do not create true asymptotic states;
they can be treated as the fields describing resonances. In our next
paper we will have to say more about this scenario. For the present we
just shut our eyes to the existence of the problem of interpretation
and consider in this paper the most general effective Hamiltonians
constructed from the infinite set of fields corresponding to free
particles with arbitrary spins and masses.

Another problem connected with Weinberg's scheme is that of
nonlinearly realized (dynamical) symmetries%
\footnote{In the case of linear (algebraic) symmetry there is no
problem at all.}. %
It is extremely difficult (if ever possible) to formulate the
conditions providing a guarantee of the desired dynamical symmetry
properties of amplitudes resulting from the effective quantum
Hamiltonian written in the interaction picture. Surely, this
difficulty is explained by the fact that neither free nor interaction
Hamiltonian by itself commutes with the dynamical symmetry generators.
The solution (at least, partial) to this problem can be obtained from
the results of Ref.
\cite{WeinEFT}.
In that paper it is shown that, leaning upon the canonical
quantization scheme and using the
`minimal'
invariant Lagrangian (that containing the minimal number of field
derivatives required by symmetry), it is possible to calculate the
lowest order terms in series expansion (in small momenta) of the
amplitude describing a process with Goldstone bosons. This means that
the dynamical symmetry requirements can be formulated -- at least, in
lowest orders -- directly in terms of amplitudes; one has no
necessity in formulating them on the Hamiltonian level%
\footnote{An example is provided by famous Low's theorems in QED.}. %
In turn, this means that, in order to account for the dynamical
symmetry requirements in the framework of effective theory, one needs
to compute amplitudes of the processes involving Goldstone bosons and
then compare the results with those obtained from the canonically
quantized invariant classical Lagrangian of the lowest order. This
very approach has been used in Refs.
\cite{VV}, \cite{AVVV}
to derive the restrictions imposed by Chiral
$SU_2 \times SU_2$
symmetry on the structure of meson resonance spectrum. The answer to
the question on how to write down the restrictions imposed by certain
kind of dynamical symmetry on the higher order amplitudes still
remains unclear. In this paper we do not discuss this point.

One note is in order. In what follows we assume that the effective
theory under consideration does not contain massless particles of
higher spin
$J > 1/2$.
This is just a technical assumption, but at the moment we do not
know how to avoid it.

\section{Classification of the parameters}
\label{sec-classification}
\mbox{}

The effective Hamiltonian contains all the types of local terms
consistent with Lorentz symmetry. For example, along with the simple
interaction term
${\phi}^4$,
it contains also the terms of the form
${\phi}^2 {\partial}_{\mu}\phi {\partial}^{\mu}\phi $,
${\phi}^2{\partial}_{\mu \nu} \phi {\partial}^{\mu \nu} \phi$,
$
\phi {\partial}_{\mu}\phi {\partial}_{\nu} \phi
{\partial}^{\mu \nu} \phi
$,
${\phi}^5$,
and so on. This means that many Hamiltonian coupling constants
contribute to the same kinematical structure in the amplitude of a
given process (say, to the term
$\sim s^2$
in the tree-level amplitude of the process
$2 \rightarrow 2$).
Hence, to perform the renormalization programme, one needs first to
solve the problem of classification of couplings in order to avoid
attracting unnecessary (dependent) counterterm vertices. Another
reason, explaining why the solution to this
{\em problem of couplings}
might happen extremely useful, is the following. As known (see, e.g.
\cite{Weinberg, Collins}),
the most difficult problem connected with renormalization of effective
theories (which are renormalizable by the very construction) is the
necessity to formulate an infinite number of renormalization
prescriptions needed to fix the finite parts of counterterms. This
looks impracticable until one finds a regularity effectively reducing
the number of independent prescriptions. It seems quite natural to
look for the mathematical expression of such a regularity in terms of
{\em independent}
parameters appearing in a theory.

Inasmuch as we are only interested in describing the scattering
processes%
\footnote{In other words, we are interested in constructing the
{\em effective scattering theory}.},
it looks reasonable to work in terms of the parameters appearing in
$S$-matrix
elements. Those parameters are the functions of Hamiltonian coupling
constants. Clearly, the Green functions of a theory depend on the same
parameters as the
$S$-matrix
elements do; but, in addition, they may depend on the
`orthogonal'
combinations only contributing off the mass shell. Hence it makes
sense to classify the parameters as
{\em essential}
and
{\em redundant}
ones (see Chapter 7.7 of Ref.
\cite{Weinberg}).
We follow the general line of this classification but we find it
necessary to make more precise definitions of the terms.

~First, we work with the quantum Hamiltonian in the interaction
picture. In contrast, the definitions in
\cite{Weinberg}
refer to the Lagrangian coupling constants. As we have already noted,
in the case of effective theory the simple connection between the
canonical Lagrangian and quantum Hamiltonian approaches happens lost
and there is no real possibility to express the Hamiltonian parameters
in terms of the Lagrangian ones. Second, in contrast with
\cite{Weinberg},
we classify the parameters appearing in the expressions for
$S$-matrix
elements of a given loop order, not only those in the Hamiltonian. The
reason for elaborating the more detailed classification of the
effective theory parameters is that the form of dependence of matrix
elements on the Hamiltonian coupling constants depends of the loop
order in question. Hence it looks quite natural to elaborate a
classification of parameters appearing in amplitudes of effective
theory at a given order of loop expansion. As shown below, it
happens possible to point out the set of independent%
\footnote{As long as coupling constants in the effective Hamiltonian
are considered independent.} %
parameters (combinations of coupling constants) quite sufficient to
describe all the
$S$-matrix elements of a given order.

Because of all these reasons we use the following definitions. The
independent combinations of Hamiltonian coupling constants needed to
fix the kinematical structure of all the renormalized
$N$-loop
Green functions of a given effective theory we call as just the
$N$-th
level parameters. We separate them into two groups. The first group
only contains those combinations of the parameters which do not appear
in the expressions for renormalized
$S$-matrix
elements. We call these combinations as the
{\em redundant parameters of the N-th level}.
All the other independent combinations we collect in the second group
and call them as the
{\em resultant parameters of the N-th level}.
The term
{\em essential}
is reserved for those combinations of resultant parameters which
appear in the
\underline{well-defined}
(converging) series presenting the amplitudes of a given loop order in
certain kinematical domains.

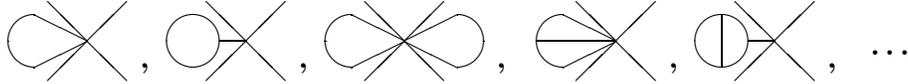
\begin{figure}[h]
\begin{center}
\begin{picture}(360,30)(-30,-15)

\put(0,-20){
\begin{picture}(60,30)(0,-15)
\put(0,0){\line(-1,1){15}}
\put(0,0){\line(-1,-1){15}}
\put(0,0){\line(1,-1){15}}
\put(0,0){\line(1,1){15}}
\put(-20,0){\oval(20,20)[l]}
\put(-20,-10){\line(2,1){20}}
\put(-20,10){\line(2,-1){20}}
\put(20,-10){\shortstack{\bf ,}}
\end{picture}
}

\put(60,-20){
\begin{picture}(60,30)(0,-15)
\put(0,0){\line(-1,1){15}}
\put(0,0){\line(-1,-1){15}}
\put(0,0){\line(1,-1){15}}
\put(0,0){\line(1,1){15}}
\put(-20,0){\circle{20}}
\put(0,0){\line(-1,0){10}}
\put(20,-10){\shortstack{\bf ,}}
\end{picture}
}

\put(120,-20){
\begin{picture}(60,30)(0,-15)
\put(0,0){\line(-1,1){15}}
\put(0,0){\line(-1,-1){15}}
\put(0,0){\line(1,-1){15}}
\put(0,0){\line(1,1){15}}
\put(-20,0){\oval(20,20)[l]}
\put(-20,-10){\line(2,1){20}}
\put(-20,10){\line(2,-1){20}}
\put(20,0){\oval(20,20)[r]}
\put(20,10){\line(-2,-1){20}}
\put(20,-10){\line(-2,1){20}}
\put(35,-10){\shortstack{\bf ,}}
\end{picture}
}

\put(200,-20){
\begin{picture}(60,30)(0,-15)
\put(0,0){\line(-1,1){15}}
\put(0,0){\line(-1,-1){15}}
\put(0,0){\line(1,-1){15}}
\put(0,0){\line(1,1){15}}
\put(-20,0){\oval(20,20)[l]}
\put(0,0){\line(-1,0){30}}
\put(-20,-10){\line(2,1){20}}
\put(-20,10){\line(2,-1){20}}
\put(20,-10){\shortstack{\bf ,}}
\end{picture}
}

\put(260,-20){
\begin{picture}(60,30)(0,-15)
\put(0,0){\line(-1,1){15}}
\put(0,0){\line(-1,-1){15}}
\put(0,0){\line(1,-1){15}}
\put(0,0){\line(1,1){15}}
\put(-20,0){\circle{20}}
\put(0,0){\line(-1,0){10}}
\put(-20,-10){\line(0,1){20}}
\put(20,-10){\shortstack{\bf ,}}
\end{picture}
}

\put(300,-10){\shortstack{\bf \ldots}}

\end{picture}
\end{center}
\caption{\label{1f} Examples of 1- and 2-loop order vertices.}
\end{figure}

The similar classification applies also for the parameters appearing
in the expression for
{\em pointlike vertex of the N-th loop order}:
that, containing self-closed lines --- bubbles or/and tadpoles ---
which, in turn, may have complex multi-loop inner structure
(see Fig.~\ref{1f}),
the total number of loops being
$N$.
The presence of an arbitrary number of such bubbles (or/and tadpoles)
does not change analytic structure of the vertex%
\footnote{It is just a polynomial or power series in kinematical
variables.},
it only may change the numerical coefficients in the corresponding
polynomials (series).

Sometimes, it is convenient to classify in the same way the parameters
appearing in the Hamiltonian. In this case we use the term
`Hamiltonian level parameters'
(effective, minimal, non-minimal).

Clearly, the full set of the parameters needed to describe the
amplitude of a given process (at a given order of loop expansion) is
exhausted by the power series expansion coefficients around an
arbitrary nonsingular point in the space of corresponding kinematical
variables. The problem is that the full collection of such sets
necessarily contains dependent parameters because general principles
(causality, crossing, etc.) impose certain limitations on its
structure. This is the reason why we work with pointlike vertices of
different loop orders and classify the coefficients appearing in
corresponding analytical expressions.

For the following it is also useful to introduce the notion of the
{\em effective vertex}.
Let us consider a formal sum of all the Hamiltonian monomials
constructed from a given set of
$n$
fields and differing from one another by the total number and/or
positions of differential operators
${\partial}_{\mu}$ (for example,
${\phi}^2 {\partial}_{\mu}\phi {\partial}^{\mu}\phi $,
${\phi}^2{\partial}_{\mu \nu} \phi {\partial}^{\mu \nu} \phi$,
$
\phi {\partial}_{\mu}\phi {\partial}_{\nu} \phi
{\partial}^{\mu \nu} \phi
$,
$\ldots$).
Each one of these monomials corresponds to an individual vertex
(polynomial in kinematical variables) in the system of Feynman rules.
It happens convenient to consider the infinite sum of all such
vertices. It takes a form of infinite formal series in powers of
variables. We call this series as the effective vertex of the
Hamiltonian order. The Hamiltonian can be rewritten in the form of an
infinite sum of effective vertices, the single items differing from
one another by the number or/and by quantum numbers of field
operators. Hence, the full sum of Feynman graphs (of a given loop
order) describing the amplitude of a given process can be always
presented as a sum of graphs written in terms of effective vertices of
the Hamiltonian order. In what follows we imply tacitly that this is
done. The problem of convergence of formal infinite sums will be
discussed in
Sec.~\ref{sec-resultant}.

The notion of the effective vertex of
$N$-th
loop order is introduced in much the same way -- this point is
considered in more detail in
Sec.~\ref{sec-minimal}.
The coefficients appearing in corresponding series we call as the
{\em N-th level effective parameters}.

By construction, the effective theory Hamiltonian contains an infinite
number of coupling constants, only a part of them (or, better, their
combinations) contributing to
$S$-matrix
elements. We do not need to compute all the Green functions of a
theory because we are only interested in the amplitudes of various
scattering processes. This means that for our purpose it is quite
sufficient to consider theories only renormalizable in the sector of
essential parameters. The divergences in Green functions unrelated to
$S$-matrix
elements%
\footnote{An excellent example of such divergences is provided
by the Standard Electroweak Model in the unitary gauge.} %
will never bother us.

Thus we need to select the set of essential parameters. This cannot be
done through just a classification of coupling constants appearing in
the Hamiltonian. The reason is that, except few trivial cases, both
essential and redundant parameters are very complicated functions of
the Hamiltonian coupling constants
$G_i$.
Suppose for a moment that all such functions are constructed and
classified as essential
($E_i$, $i=1,2,...$)
and redundant
($R_i$, $i=1,2,...$)
parameters. This would provide us with an infinite system of algebraic
equations of the form
\begin{equation}
E_i = E_i(G_1,...), \ \ \ \ \ \ \ \ \ \ R_j = R_j(G_1,...) \ \ \ \ \ \
\ \ \ \ \ \
(i,j = 1,2, \ldots)\ \ .
\label{6.1a}
\end{equation}
Resolving this system with respect to
$G_i$,
one obtains
\begin{equation}
G_i = G_i(E_1,...,R_1,...) \ \ \ \ \ \ \ \ \ \
(i = 1,2, \ldots)\ \ .
\label{6.1b}
\end{equation}
Hence, when dealing with
$S$-matrix
elements, one can assign to
$R_i$
whatever values convenient for computations. In particular, there is
no necessity in formulating the renormalization prescriptions fixing
the finite parts of
`redundant counterterms'.
In turn, this is especially useful if we are interested in finding a
regularity allowing to put in order the infinite system of
normalization prescriptions needed to compute amplitudes of various
scattering processes in the framework of effective theory.

Thus we see that it would be extremely useful if we find a way to
write down the explicit form of the relations
(\ref{6.1a}).
We do not know how to solve this problem in general. Instead, one can
try to find a
{\em perturbative}
solution providing the required relations at every fixed order of loop
expansion. To realize this idea, one needs to perform certain
reconstruction (reparametrization) of the initial Dyson series. Below
we describe a special kind of parametrization which serves this
purpose.

We imply tacitly that there exists a regularization consistent with
all the desired symmetries. This suggestion looks harmless if the
Euclidean version of a theory is considered. However, if one works in
Minkowski space (as we do), it seems much less trivial. Nevertheless,
we believe that it is true.

This classification happens especially convenient when one is only
interested in computing the renormalized
$S$-matrix
elements in the effective theory framework. As far as we know, the
special role of essential parameters has been first stressed in
\cite{WeinAsySafe}.

The logical line of subsequent consideration is the following. We
start from the basic effective Hamiltonian written in terms of the
`physical' masses and `physical' couplings%
\footnote{When speaking about the mass of unstable particle, it is more
appropriate to use the term `renormalized'. In fact, this implies
using the renormalized perturbation theory with the conventional OMS
(on-mass-shell) normalization conditions (see, e.g.
\cite{Collins} --
\cite{Tkachov}).
The quotation marks are used to stress that only certain combinations
of the Hamiltonian parameters present measurable quantities.}
(plus counterterms). Next, we note that some combinations of
$G_i$
(their forms depend on the loop order in question) certainly
contribute to measurable quantities and thus could be considered, at
least, as building blocks for essential parameters. Then we prove that
it is always possible to rewrite the expression for arbitrary graph of
a given loop order in such a way that the renormalized
$S$-matrix
only depends on those latter combinations called below as
{\em minimal parameters}.
Further, we show that the full sum of graphs of the same loop order
can be rewritten in a form quite similar to that constructed on the
previous step for an individual graph with a fixed set of internal
lines. The parameters appearing on this stage are called as the
{\em resultant parameters}.
At last, we direct the way allowing to construct all the essential
parameters as certain infinite sums of the resultant ones. This result
shows that, when dealing with effective scattering theory, it is
always possible (at least, in principle) to make use of the scheme of
renormalized perturbation theory only appropriate in the sector of
essential parameters.

\section{Minimal parameters}
\label{sec-minimal}
\mbox{}

The immediate task we are going to solve is to prove the following
statement.
{\em The full set of the essential parameters of effective theory is
constructed solely from those combinations of the Hamiltonian coupling
constants (including masses) which are needed to fix the independent
on-shell kinematic structures appearing in the expressions for
effective vertices (of different orders) multiplied by the
relevant wave functions}. %
In fact, this statement is almost trivial but its precise meaning
deserves comments. This Section is devoted to the preliminary
consideration needed for better understanding of the proof given in
Sec.~\ref{sec-proof}.

The proof consists of two steps. First, we show that all invariant
structures (formfactors), describing a given vertex in arbitrary
$S$-matrix
graph, can be reduced to a simpler form (called as minimal). Second,
we show that it is always possible to reduce the full set of tensor
structures needed to fix the form of this vertex to a subset only
containing a part of them (also called as minimal). The corresponding
procedure -- called below as the graph reduction -- eliminates certain
part of the parameters which we call as non-minimal. When applied to a
given graph, it results in the sum of two items. The first one is just
the initial graph written in terms of new -- minimal -- vertices (of
different orders) completely described by the relevant minimal
parameters. The second item does not contribute to the renormalized
$S$-matrix
under the condition that the normalization point is taken on mass
shell.

We would like to note that the reduction procedure is only needed to
prove the completeness of the full set of minimal parameters. We do
not imply its using in practical calculations.

It is a point here to stress the difference between two terms often
used throughout the paper. The term
{\em on-shell graph}
means that the graph in question (say, pointlike vertex) is computed
at all external momenta on the mass shell. The term
{\em S-matrix graph}
(or, the same,
{\em amplitude graph})
means that the on-shell graph is dotted by the relevant wave
functions. The difference between the corresponding expressions
manifests itself in the case when particles with spin
$J \neq 0$
are considered.

Now we need to explain the precise meaning of the term
{\em minimal}
(minimal vertex, minimal propagator). The reason why we use one more
special term in addition to those defined above (essential, redundant)
is explained by the following circumstance. The difference between the
essential and redundant parameters manifests itself when one considers
the structure of the amplitude of a given scattering process. This
amplitude results from contributions of many different
$S$-matrix
graphs of a given loop order. Thus the essential parameters of a given
level happen constructed from the Hamiltonian coupling constants
describing the vertices with different numbers of field operators.
This language is not suitable for discussing the problems of
renormalization. That is why we need the more detailed classification
of various combinations of the Hamiltonian coupling constants
appearing in the process of calculation of a given graph.

Consider an effective vertex
$V_{...}(p_1, \ldots , p_n)$
(the ellipses stand for Lorentz indices) with
$n$
lines carrying the momenta
$(p_1, p_2, \ldots, p_n)$
only restricted by the conservation law. As explained in the previous
Section, this vertex corresponds to an infinite sum of monomials in
the Hamiltonian. Each monomial is constructed from fields and their
derivatives, the total number of field operators
being
$n$.
The explicit expression for this vertex reads
\begin{equation}
V_{...}(p_1, \ldots , p_n) = \sum \limits _{a=1}^{M+N}
                              T_{...}^{(a)} F_a\ ,
\label{6.1}
\end{equation}
where
$T_{...}^{(a)}$
stand for whatever independent tensor structures needed
(their total number is denoted as
$M+N$)
and
$F_a$ --
for the corresponding scalar formfactors (formal power series in
invariant kinematical variables).

It is pertinent to remind that the expression
(\ref{6.1})
is equally applicable in the case if we consider the pointlike
vertex of the
$L$-th
order. In accordance with the definition given in
Sec.~\ref{sec-classification},
the corresponding coefficients of formal power series for
$F_a$
are called as the
$L$-th
level parameters.

Further, choose a set of independent scalar variables (their
total number is
$4n-10$)
as follows
\begin{equation}
\left[ {\pi}_1, \ldots , {\pi}_n;
       {\nu}_1, \ldots , {\nu}_{3n-10}
\right].
\label{6.1c}
\end{equation}
Here
$$
{\pi}_i \equiv p_i^2 - m_i^2\ ,
$$
and
${\nu}_r$
stand for the rest (arbitrarily choosen%
\footnote{The problem of appropriate choice of those
variables will be discussed in more detail in a separate
publication.})
independent linear combinations of scalar products
\begin{equation}
{\nu}_r \equiv \sum \limits_{i,j = 1}^{n}
s_{ij}^r (p_i \cdot p_j)\ , \ \ \ \ \ \ \ \ \ \
(r = 1, \ldots, 3n-10)
\label{6.2}
\end{equation}
with numerical coefficients
$s_{ij}^r$.

It is always possible to rewrite
$F_a$
as follows
$$
F_a({\pi}_1,...,{\pi}_n;{\nu}_1, ..., {\nu}_{3n-10}) =
F_a({\pi}_1,..., {\pi}_{i-1},0,{\pi}_{i+1},...,{\pi}_n;
{\nu}_1, ..., {\nu}_{3n-10}) + {\pi}_i P_a(...)\ .
$$
Thus the vertex under consideration takes a form of a sum of two
items:
\begin{equation}
V_{...}(p_1, \ldots , p_n) =
\sum \limits _{a}^{}T_{...}^{(a)} \left[
F_a^{(i)} + {\pi}_i P_a \right]\ .
\label{6.2a}
\end{equation}
The scalar functions
$F_a^{(i)}$
appearing in the first term are called as minimal with respect to the
$i$-th
line. They do not change their form when this line is put on its mass
shell. The second -- non-minimal -- term vanishes in this case.

We call the propagator as minimal if its numerator is just a spin sum
written in a covariant form and considered as a function of four
independent variables
$p_{\mu}$.
The non-minimal propagator differs from the minimal one by
non-pole terms%
\footnote{It is this point where our suggestion on the absence
of massless particles of higher spin happens important.}. %
In what follows we imply using the minimal propagators. This does not
reduce the generality of our analysis because non-pole terms result in
precisely the same effect as that caused by non-minimal parameters.
This will become more clear after reading the next
Section. Besides, as shown in
\cite{VV}, \cite{AVVV},
in practical calculations in the framework of the Cauchy form
techniques one only needs to know the residues of propagators.

Next, let us consider the tensor structures
$T_{...}^{(a)}$
occurring in
(\ref{6.1}).
They may contain the factors
($\gamma$
matrices, tensors
$g_{\mu \nu}$
and
${\varepsilon}_{\alpha \beta \gamma \delta}$,
momentum
$p_i^{\mu}$)
resulting in constants when the line in question is put on the mass
shell and dotted by the corresponding wave function. We call such
factors as non-minimal (with respect to a given line). For example, if
the line under consideration corresponds to a vector particle (with
momentum
$p_i$),
every tensor structure containing
$p^{\mu}_i$
is classified as non-minimal.

The full set of independent tensor structures
$T_{...}^{(a)}$
can be separated into two groups as follows%
\footnote{The numbers
$M_i$ and $N_i$
depend on spin of the line in question; the total number
$M_i + N_i$
of tensor structures only depends on the vertex type.}:
\begin{equation}
T_{...}^{(a)} =
\left\{
T_{...}^{(1,i)}, \ldots , T_{...}^{(M_i,i)};
R_{...}^{(1,i)}, \ldots , R_{...}^{(N_i,i)}
\right\}\ \ \ \ \ \ \ \ \ \ \
\label{6.3}
\end{equation}
Here, the first group
\begin{equation}
T_{...}^{(k,i)} \ \ \ \ \ \ \ \ \ \
(k=1,...,M_i;\ \ 1 \leq i \leq n)
\label{6.4}
\end{equation}
does not contain any non-minimal (with respect to
$i$-th
line!) structures, while the second one
\begin{equation}
R_{...}^{(k,i)} \ \ \ \ \ \ \ \ \ \
(k=1,...,N_i;\ \ 1 \leq i \leq n)
\label{6.5}
\end{equation}
consists of all such structures. The structures from the first group
are called as minimal (with respect to the given line). The meaning of
this separation is explained by the fact that, when dotted by the
relevant propagator or wave function, the non-minimal structures
result in the same terms as the minimal ones or/and in terms
proportional to
${\pi}_i$.
In other words, the effect of non-minimal tensor structures is quite
similar to that of non-minimal parameters appearing in scalar
formfactors.

By way of illustration, let us consider the case of non-minimal
structure containing the factor
$p_{\mu}$
corresponding to a vector particle (with 4-momentum
$p$).
If the line in question is external, this structure does not
contribute to
$S$-matrix
due to the transversality of the vector particle wave function. In the
case of internal line this factor is multiplied by the vector particle
propagator. The resulting expression does not contain a pole:
$$
p_{\mu}
\frac{-{g^{\mu}}_{\nu} + p^{\mu} p_{\nu}/M^2}{p^2 - M^2}
= \frac{1}{M^2} p_{\nu}\ .
$$
This means that inside a graph the non-minimal structure plays a role
of
`pole killer'.

Note, that non-minimal structures never survive as independent items
in the expressions for scattering amplitudes.

The vertex is called as minimal if it is minimal with respect to all
its lines and the corresponding expression does not contain any
non-minimal tensor structures. The algebraic form of Lorentz invariant
expression for minimal vertex does not change its appearance when the
momenta are considered on the mass shell
${\pi}_i = 0\ (i=1,\ldots, n)$.

The explicit form of the minimal vertex differs from that of
non-minimal one by the items proportional to
$(p_i^2 - m_i^2)$
or/and by those proportional to non-minimal (at least, with respect to
one of the lines) tensor structures%
\footnote{This is also true with respect to the vertices containing
self-closed lines (`bubbles' or `tadpoles'). As explained in
Sec.~\ref{sec-classification},
we classify such vertices as pointlike.}. %
Inside a graph such terms work as
`pole killers'.
This very property provides a basis for the statement formulated in
the beginning of this Section.

The instructive example where the difference between minimal and
non-minimal elements (vertices and propagators) manifests itself
explicitly (and happens important) is provided by the conventionally
used propagator and interaction Hamiltonian of the spin-3/2
Rarita-Schwinger field (see, e.g.
\cite{Ellis1}, \cite{Ellis2}
and references therein). This field corresponds, in particular,
to the well established resonance
$\Delta (1232)$
playing an important role in low energy pion-nucleon processes.
Because of this reason this field is often used in various Lagrangian
models. The problem appears when different authors use different forms
of the interaction term and (or) propagator, this difference sometimes
leading to contradictive results%
\footnote{For the references and discussion see, e.g.
\cite{UFN}, \cite{Analysis}.}.
The most popular forms of those elements used in the literature differ
from one another by the terms resulting in a
`pole killer'.
This difference produces an additional (smooth) contribution to the
amplitude which, in turn, changes the results of data fitting. This is
just an artifact of Rarita-Schwinger
formalism%
\footnote{This would not occur if Weinberg's formalism
\cite{WeinQuant}
for spin-J field were used.}.
Surely, the pole term happens the same in both cases, the residue
being just a spin sum. The so-called off-shell couplings turn out
to be redundant (see
\cite{Ellis1}, \cite{Analysis}).

\section{The proof of the statement}
\label{sec-proof}
\mbox{}

To prove the statement formulated in the beginning of the previous
Section, it is sufficient to show that
{\em an arbitrary
$S$-matrix
graph can be rewritten in the form only constructed from the minimal
vertices of different orders plus the terms which do not contribute to
renormalized S-matrix}.

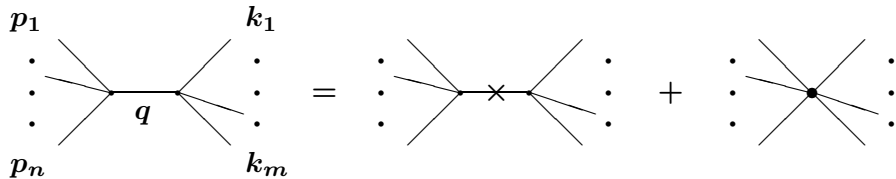
\begin{figure}[th]
\begin{center}
\begin{picture}(353,50)(25,80)

\put(75,100){\line(1,0){25}}             
\put(80,90){ \shortstack{\boldmath$q$} } %

\put(75,100){\circle*{2}}
\put(75,100){\line(-1,1){20}}
\put(75,100){\line(-1,-1){20}}
\put(75,100){\line(-4,1){25}}

\put(33,125){ \shortstack{\boldmath$p_1$} } 
\put(33,70){ \shortstack{\boldmath$p_n$} }  %

\put(45,88){\circle*{2}}    %
\put(45,100){\circle*{2}}   
\put(45,112){\circle*{2}}   %

\put(100,100){\circle*{2}}
\put(100,100){\line(1,1){20}}
\put(100,100){\line(1,-1){20}}
\put(100,100){\line(3,-1){25}}

\put(122,125){ \shortstack{\boldmath$k_1$} } 
\put(122,70){ \shortstack{\boldmath$k_m$} }  %

\put(130,88){\circle*{2}}    %
\put(130,100){\circle*{2}}   
\put(130,112){\circle*{2}}   %

\put(147,97){ \shortstack{\boldmath$=$} } 


\put(207,100){\line(1,0){25}}                  
\put(212,97){ \shortstack{\boldmath$\times$} } 

\put(207,100){\circle*{2}}
\put(207,100){\line(-1,1){20}}
\put(207,100){\line(-1,-1){20}}
\put(207,100){\line(-4,1){25}}

\put(177,88){\circle*{2}}    %
\put(177,100){\circle*{2}}   
\put(177,112){\circle*{2}}   %

\put(233,100){\circle*{2}}
\put(233,100){\line(1,1){20}}
\put(233,100){\line(1,-1){20}}
\put(233,100){\line(3,-1){25}}

\put(263,88){\circle*{2}}    %
\put(263,100){\circle*{2}}   
\put(263,112){\circle*{2}}   %

\put(278,97){ \shortstack{\boldmath$+$} } 

\put(340,100){\circle*{4}}
\put(340,100){\line(-1,1){20}}
\put(340,100){\line(-1,-1){20}}
\put(340,100){\line(-4,1){25}}
\put(340,100){\line(1,1){20}}
\put(340,100){\line(1,-1){20}}
\put(340,100){\line(3,-1){25}}

\put(310,88){\circle*{2}}    %
\put(310,100){\circle*{2}}   
\put(310,112){\circle*{2}}   %

\put(370,88){\circle*{2}}    %
\put(370,100){\circle*{2}}   
\put(370,112){\circle*{2}}   %

\end{picture}
\end{center}
\caption{\label{2f} Line reduction procedure}
\end{figure}

The proof is straightforward. Consider an arbitrary complex graph%
\footnote{Regularization is tacitly implied.}
(amputated Green function) constructed in accordance with Feynman
rules derived from the effective theory Ha\-mil\-to\-ni\-an%
\footnote{It is important that we consider a graph constructed from a
fixed set of effective vertices of the Hamiltonian order; no summation
over the different types of inner lines as well as over different
types of the effective vertices is implied on this stage.}. %
~Further, consider the inner line
$q$
connecting the vertices
${V_1}^{\mu \ldots} (p_1, \ldots , p_n, q)$
and
${V_2}^{\nu \ldots} (k_1, \ldots , k_m, q)$
(see Fig.~\ref{2f}).
We do not make any suggestions about the other lines: a part of
them may be taken external while the rest ones -- internal.

First, let us consider the case when this line is the only one
connecting the vertices in question and the propagator contains
non-minimal terms:
$$
{\cal P}^{\cdots}_{,,,}(q) =
{1\over q^2 - m^2}
\left(
\Pi^{\cdots }_{,,,}(q)
+ \underbrace{
(q^2 - m^2) \phi^{\cdots }_{,,,}(q)
}_{\rm non-minimal \; terms}
\right)\; .
$$
Here
$\phi^{\cdots}_{,,,}(q)$
is some nonsingular tensor and
$\Pi^{\cdots }_{,,,}(q)$
is just a spin sum written in a covariant form and understood as a
function of four
{\em independent}
components of momentum. Besides, let us write down the vertices in the
form
(\ref{6.2a})
explicitly showing the presence of non-minimal terms in scalar
formfactors:
$$
V_{1,2}^{\cdots}(\ldots, q) =
\sum \limits _{a}^{}T^{(a)\cdots} \left[
F_a^{1,2} + (q^2-m^2) P_a^{1,2} \right]
$$
(in what follows we omit tensor indices).

It is easy to understand that non-minimal terms just kill the
denominator of the propagator and thus result in a new quasi-vertex
with
$(n+m)$
lines
$(p_1, \ldots , p_n, k_1, \ldots , k_m)$.
In other words, one can represent (rewrite) the graph in the
following way (below
$\delta (\ldots)$
denotes the momentum conservation delta-function needed for each
vertex):
$$
\ldots
\int d q \delta(\ldots) \delta(\ldots)
V_1 {\cal P} V_2
=
\ldots
\int d q \delta(\ldots) \delta(\ldots)
\left(
\sum \limits _{a}^{}T^{(a)} F_a^1
\right)
{\Pi(q) \over  q^2 - m^2}
\left(
\sum \limits _{b}^{}T^{(b)} F_b^2
\right)
$$
$$
+
\ldots
\underbrace{
\delta ({\scriptstyle \sum} p_i - {\scriptstyle \sum} k_i)
\int d q \delta ({\scriptstyle \sum} p_i - q)
\ldots
}_{\rm new\; vertex}
\; ,
$$
where ellipses before the integral stand for the rest part of
the graph. Besides, the minimal elements of
$V_1$ and $V_2$
transform the line
$q$
into a new one
(`minimal',
labelled by a cross). Thus, the initial graph gets transformed into
two new ones. The first graph has the same structure as the initial
one except that the forms of the vertices
$V_1$ and $V_2$
have been changed --- the terms proportional to
$q^2 - m^2$
disappeared and the minimal propagator appeared in place of
non-minimal one. The second graph has quite a different structure: the
new pointlike quasi-vertex with
$(n+m)$
lines has appeared in place of two original ones ---
$V_1$ and $V_2$.
This quasi-vertex does not follow from the Feynman rules based on the
effective Hamiltonian. Nevertheless, it has precisely the same
analytic structure as that of
``true''
vertex with the same number of lines. The only difference is that the
crossing symmetry properties may happen broken if the initial graph
was not properly symmetrized with respect to the lines under
consideration. Clearly, this difficulty would never appear if --- in
place of single graph --- we consider the symmetric sum of all its
topological copies. Below we imply that this is the case. This means
that the effect of non-minimal terms results in a symmetric sum of
corresponding quasi-vertices. We call this sum as the
{\em secondary vertex of order zero}
(or, the same, tree order secondary vertex). Recall, that we are
dealing with an effective theory, hence all possible vertices are
already included. Thus our procedure (later on we call it as the
reduction of a given line) only leads to a renormalization of the
parameters fixing the form of the Hamiltonian order effective vertex
with
$(n+m)$
lines.

\begin{figure}[th]
\begin{center}
\begin{picture}(390,50)(0,30)

\put(20,50){\oval(80,60)[r]} 
\put(80,50){\oval(80,60)[l]} %
\put(27,40){\shortstack{\boldmath$q_1$}}
\put(63,40){\shortstack{\boldmath$q_2$}}

\put(82,48){\shortstack{\boldmath$=$}}

\put(90,50){\oval(80,60)[r]}                 
\put(105,48){\shortstack{\boldmath$\times$}} %
\put(150,50){\oval(80,60)[l]}                %
\put(125,48){\shortstack{\boldmath$\times$}} %
\put(97,40){\shortstack{\boldmath$q_1$}}
\put(133,40){\shortstack{\boldmath$q_2$}}

\put(150,48){\shortstack{\boldmath$+$}}


\put(205,50){\line(-1,1){30}}
\put(205,50){\line(-1,-1){30}}
\put(205,50){\line(1,-1){30}}
\put(205,50){\line(1,1){30}}
\put(185,50){\oval(20,20)[l]}
\put(165,40){\shortstack{\boldmath$q_1$}}
\put(185,40){\line(2,1){20}}
\put(185,60){\line(2,-1){20}}
\put(170,48){\shortstack{\boldmath$\times$}}

\put(235,48){\shortstack{\boldmath$+$}}

\put(275,50){\line(-1,1){30}}
\put(275,50){\line(-1,-1){30}}
\put(275,50){\line(1,-1){30}}
\put(275,50){\line(1,1){30}}
\put(295,50){\oval(20,20)[r]}
\put(307,40){\shortstack{\boldmath$q_2$}}
\put(295,40){\line(-2,1){20}}
\put(295,60){\line(-2,-1){20}}
\put(300,48){\shortstack{\boldmath$\times$}}

\put(320,48){\shortstack{\boldmath$+$}}

\put(360,50){\line(-1,1){30}}
\put(360,50){\line(-1,-1){30}}
\put(360,50){\line(1,-1){30}}
\put(360,50){\line(1,1){30}}

\end{picture}
\end{center}
\caption{\label{3f} Example of reduction of two adjacent lines}
\end{figure}
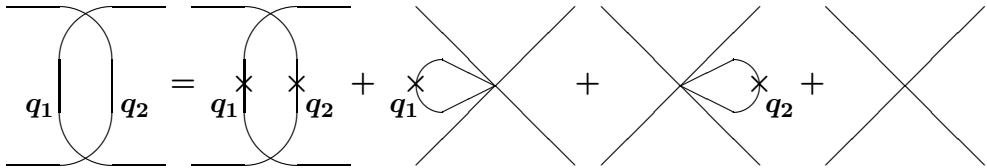

The case when there are two lines
($q_1$
and
$q_2$)
connecting the vertices under consideration can be analyzed precisely
in the same way as above. The result is illustrated in
Fig.~\ref{3f}
(for simplicity, here
$V_1$ and
$V_2$ are taken to be four-vertices).

So, in this case the reduction of both lines results in a sum of
two kinds of graphs (see
Fig.~\ref{3f}):
\begin{enumerate}
\item
The same graph as the initial one but with two crossed (minimal)
lines in place of two original ones.
\item
Three graphs with pointlike vertices dotted by the factors stemming
from crossed or uncrossed self-closed lines and from the effect of
`pole killers'%
\footnote{When the self-closed line corresponds to a particle with
spin
$J \neq 0$
these factors may result in additional reparametrization.}. %
Purely for the sake of uniformity, one can further rewrite the graph
with uncrossed bubble as a sum of two items: the same graph as the
initial one but with the crossed bubble in place of uncrossed one plus
the reminder caused by the effect of relevant
`pole killers'.
We would like to stress that each one of these pointlike graphs should
be considered as the 1-loop order graph whether or not the bubble is
drawn explicitly (see the last of graphs shown in
Fig.~\ref{3f}).
\end{enumerate}

Proceeding in the same way one can realize that, in the case when
there are
$l$
lines connecting
$V_1$
and
$V_2$,
the reduction procedure creates the same two vertices with
$l$
minimal lines in place of the original ones. Besides, it creates a set
of vertices with more (also minimal) external lines and
$n < l$
bubbles (visible or/and invisible) some of which, in turn, may present
a complex loop structure. It is important that these new vertices
possess the pointlike kinematical structure.

In order to preserve the loop counting rules, we use special term for
the sum of secondary quasi-vertices resulting from the reduction of
one of
$l$
lines connecting two vertices under consideration. This sum can be
considered as a single
{\em secondary vertex of the
$(l-1)$-th
order}.
In general, the pointlike vertex with several bubbles (tadpoles),
having in total
$L$-loops,
is called as the secondary vertex of the
$L$-th order.
For example, the sum of three pointlike graphs depicted in
Fig.~(\ref{3f})
is defined as a single secondary vertex of the first order.

When continued for all internal lines of a given graph, the reduction
procedure results in a sum of graphs constructed from minimal
propagators and pointlike vertices (with different number of bubbles)
in which all the lines are minimal except those happened to be
external in the initial graph. But each one of these latter lines may
happen internal in the case if a given vertex appears also in the
inner part of the graph! To avoid inconsistency, let us present the
vertices connected with external lines in the form
(\ref{6.2a}):
$$
V_{...}(p_1, \ldots , p_n) =
\sum \limits _{a}^{}T_{...}^{(a)} \left[
F_a^{(i)} + {\pi}_i P_a \right]\ .
$$
This results in a sum of graphs which can be divided into two groups.
The first group consists of all graphs constructed solely in terms of
minimal propagators and minimal vertices of different orders. In
contrast, every graph from the second group contains at least one
vertex of the type
$P_a$
connected with one of the external lines (say,
$i$-th)
and dotted by the factor
${\pi}_i$
corresponding to this line.

Graphs from the second group (let us call them as non-minimal) do not
contribute to the amplitude of the process under consideration.
Nevertheless, they cannot be simply neglected. The point is that those
graphs might result in nontrivial contributions of two different
kinds. First, they contribute to the amplitudes (of the same loop
order as that in question) corresponding to the processes involving
more particles. Second, they can contribute to the values of
renormalization constants.

The contribution of non-minimal graphs (with a given number of
external lines) to the amplitudes of the processes involving more
particles can be rewritten in terms of minimal parameters precisely in
the same way as above. In what follows we tacitly imply that this is
done with respect to all
$S$-matrix
graphs of the loop order
$L$
under consideration.

As to the influence of non-minimal graphs on the values of
renormalization constants, it happens irrelevant if we are only
interested in the corresponding
$S$-matrix
elements of a given order and, in addition, rely on the conventional
OMS renormalization scheme
(see, e.g.
\cite{Nekrasov}, \cite{Sirlin}).
In this case one can simply forget about this group of graphs because
the only quantities depending on their parameters are the wave
function renormalization constants%
\footnote{This relates to the case of self-energy graphs.}
which, in turn, are just redundant parameters having no influence
on renormalized
$S$-matrix
elements of the order in question. In the opposite case, when one
needs to calculate the amplitudes of the loop order
$(L+1)$,
the
$L$-th
order non-minimal graphs cannot be neglected. Instead, they must be
taken into account when constructing the next order graphs which, in
turn, should be further subjected to the reduction procedure. The
important point is that, after this is done, the parameters appearing
in non-minimal graphs of the
$L$-th
loop order will happen absorbed into the structure of minimal
parameters describing the vertices of the order
$(L+1)$,
this being also true with respect to
$L$-th
order non-minimal counterterms. From this note it follows the
important conclusion:
{\em to obtain finite results for
$S$-matrix
elements in the framework of effective theory, one has no need in
formulating the normalization conditions fixing the finite parts of
non-minimal counterterms}.

Now, the first step is done. We have shown, that it is always possible
to pick out certain group of parameters which do not produce the
kinematically independent contributions to renormalized amplitudes at
a given order of loop expansion. So, from this point we can consider
the scalar formfactors
$F_a$
being minimal with respect to each line. This, in turn, means that
they only depend on kinematical variables
(\ref{6.2}),
the dependence on
$\pi_i$
may be dropped.

We would like to stress once more that the above analysis is only true
in the framework of OMS renormalization scheme: the renormalization
point must be taken on mass shell. It is this condition which allowed
us to consider both external and internal lines on the same footing.
In turn, this means that for unstable particles the Hamiltonian mass
parameters may happen only indirectly connected with pole positions of
the corresponding full propagators
(see
\cite{Nekrasov}, \cite{Sirlin}).

Thus in order to calculate the amplitude of a given scattering process
up to a given order of loop expansion, one only needs to formulate the
normalization prescriptions for the remaining group of parameters.
However, as yet this cannot be done in terms of measurable quantities
because this latter group still contains the redundant combinations.
To reveal them we need to consider the influence of non-minimal tensor
structures.

Let us rewrite each of the vertices
$V_1$, $V_2$
as follows
\begin{equation}
V_{...}(p_1, \ldots , p_n) =
\sum \limits _{a=1}^{M} T_{...}^{(a)} F_a^t +
\sum \limits _{a=1}^{N} R_{...}^{(a)} F_a^r\ .
\label{7.1}
\end{equation}
The first sum in
(\ref{7.1})
contains all the independent minimal (with respect to each of the
lines!) tensor structures
$
T_{...}^{(a)},
$
while the second one contains all the other independent structures
(non-minimal, at least, with respect to one of the lines). This means
that every coefficient of the polynomials (series)
$$
F_a^t({\nu}_1,...,{\nu}_{3n-10})
$$
presents a measurable quantity%
\footnote{Strictly speaking, this is not quite true. It would be
better to say that those coefficients contribute to measurable
quantities. The point is that it is impossible to measure the
contribution of the individual vertex -- only a sum of all the
relevant graphs of a given order presents the measurable quantity.
We will come back to this point below.}. %
This is so just because each one of those coefficients results in
the individual kinematical structure in the amplitude.

Hence we conclude that all the combinations of coupling constants
appearing (as expansion coefficients) in the invariant formfactors
$F_a^{t}$
should be classified as building blocks for the essential parameters.

Now, let us consider the parameters from the second group, namely,
those appearing in the formfactors
$F_a^{r}({\nu}_1,...,{\nu}_{3n-10})$
describing the contributions of non-minimal tensor structures
$R_{...}^{(a)}$.
Below it is shown that the effect produced by this group is reduced to
just a renormalization of the minimal parameters (those appearing in
``minimal''
formfactors
$F_a^t({\nu}_1,...,{\nu}_{3n-10})$).

For simplicity, we only consider here the case of structures of the
bosonic type. The generalization for fermions is straightforward. To
describe the fields with spin
$J \neq 0$
we use the conventional Rarita-Schwinger formalism
\cite{Rarita}
and rely upon the method of contracted projecting operators
(see, e.g.
\cite{WeinQuant}, \cite{Scadron}, \cite{Alfaro}).
The corresponding wave functions
${\epsilon}_{{\mu}_1...{\mu}_J}(j,q)$
(here
$q$
stands for momentum and
$j$ --
for polarization) posses symmetry, tracelessness and transversality
properties.

~First, consider the case when one of the lines of the vertex (say,
$V_1$)
is external and corresponds to a particle with spin
$J \neq 0 $
and momentum
$p$.
We are only interested in non-minimal tensor structures, hence the
relevant expression necessarily contains the terms of the form
$$
p_{{\mu}_1}...p_{{\mu}_J},\ \ \ \ \ \ g_{{\mu}_1 {\mu}_2}
p_{{\mu}_3},...,p_{{\mu}_J},\ \ \ \ \ \
\ldots .
$$
The corresponding amplitude graph equals zero.

Now, consider the case when this line is internal. Keeping in mind
that the numerator of the propagator is just a spin sum (written in
covariant form and considered as a function of four independent
components of momentum), it is easy to understand that non-minimal
tensor structures result in polynomial contributions. This follows
from the fact that in this case the residue equals zero due to the
properties of spin sums. The symmetry, tracelessness and
transversality%
\footnote{Plus
$\gamma$-transversality
in the case of fermion fields.} %
properties (only valid on the mass shell!) play precisely the same
role as
`pole killers'
discussed above. Thus we conclude that the only effect produced by
non-minimal tensor structures is reduced to a renormalization of the
coefficients in invariant formfactors
$F_a^{t}$.
This may result in reappearing of the variables
${\pi}_i$
but now we know how to manage this problem: it is sufficient to
repeat the reduction procedure once more.

Thus it is shown that, at every fixed order
$L$
of loop expansion, the contribution of an arbitrary graph to the
amplitude of a given process can be rewritten solely in terms of
minimal parameters of the
$L$-th
and lower levels. Hence, all the essential parameters of the
$L$-th
level are constructed solely from those minimal parameters. Note, that
no distinction between the basic and counterterm vertices has been
made in the course of our analysis.

In particular, this means that, when calculating the amplitude of
pion-nucleon scattering in a framework of effective theory, one
can use the
`non-chiral'
interaction Hamiltonian
$\bar{N} {\gamma}_5 N \pi$:
this does not necessarily lead to a contradiction with chiral
invariance.

We will say that the amplitude graph%
\footnote{Recall that the proper symmetrization with respect to
all the lines of identical particles is tacitly implied.}
of a given (true!) loop order
$L$
is presented in the
{\em minimal (or, unitary) parametrization},
if it is rewritten in terms of minimal propagators and minimal
vertices of different orders
$l \leq L$.
The graph constructed solely from minimal elements we call as the
{\em minimal graph}.

As follows from the above analysis, the reduction procedure
transforms a given
$L$-loop
graph, constructed in accordance with conventional Feynman rules, into
a sum of minimal graphs of different topological structure plus the
sum of graphs with at least one non-minimal external line. When
drawing the minimal graphs, it is convenient to supply every vertex
$V_i$
with the special index
$l_i$
showing its order. The value
$l=0$
should be assigned to all the initial Hamiltonian vertices as well
as to the secondary vertices of the tree level%
\footnote{It should be kept in mind that there is no difference
between the Hamiltonian and tree levels in the case of triple
vertices.}. Under this condition, the
`true'
loop order
$L$
of the minimal graph with
$L_{min}$
loops and
$p$
vertices
$V_1,...,V_p$
of orders
$l_1,...,l_p$
equals
$$
L = L_{min} + \sum\limits_{i=1}^{p} l_i\ .
$$
The corresponding counterterm vertex
$V_c$
should be supplied with the index
$l_c = L$.
The important point is that,
{\em as far as we consider all the Hamiltonian couplings as independent
constants, the minimal parameters describing vertices of different
orders are also independent}. This statement can be easily proved by
induction.

The special convenience of dealing with minimal parametrization
becomes clear from the following note. In the cases of customary
finite-component renormalizable theories (as well as their
infinite-component
`vector copies')
one needs to formulate as many normalization prescriptions as there
are coupling constants (including masses) in the basic Hamiltonian.
This is so because of two reasons. First, in those cases we are
interested in complete renormalizability of a theory; this means that
we have to fix the finite parts of all the counterterms including
those needed to renormalize the off-shell Green functions. Second, in
conventional renormalizable theories every coupling constant presents
an essential parameter%
\footnote{Recall, that the gauge fixing parameter in gauge theories
appears in the framework of Lagrangian formalism.}. %
The situation looks much more complicated in effective theories. In
this case there are certain combinations of the Hamiltonian parameters
which do not contribute to renormalized
$S$-matrix
and, hence, cannot be related to any observable. In fact, these --
redundant -- combinations are not needed at all if we are only
interested in describing scattering processes. The reason why the
minimal parameterization happens most suitable in the case of
effective scattering theory, is that it provides us with the
(infinite) set of constants needed to construct the full set of the
essential parameters directly connected with observable quantities.
The structure of this connection is discussed in more detail in
Sec.~\ref{sec-essential}.
However, before discussing this structure we need to introduce the
notion of resultant parameters.

\section{Resultant parameters}
\label{sec-resultant}
\mbox{}

In the previous Section we have considered an individual amplitude
graph (more precisely, a symmetric sum) with a given number of
external lines and certain fixed set of inner vertices. However, this
graph (sum) only presents a part of the
$L$-th
order contribution to the amplitude describing the process under
consideration. To obtain the net result, one needs to make four steps
more.
\begin{enumerate}
\item
~First, it is necessary to carry out the reduction of all the graphs
(of the order
$L$)
with the same set of external lines but with different structure of
the set of vertices (different numbers of legs), no summation over the
kinds of internal lines (virtual particles) being implied on this
step.
\item
Second, it is necessary to sum over all possible kinds of inner
lines in every graph considered above.
\item
Third, it is necessary to sum up all the expressions obtained on the
previous steps.
\item
Fourth, it is necessary to take account of contributions due to
counterterm  vertices of the
$L$-th
loop order.
\end{enumerate}

The same should be done with respect to all the amplitude graphs with
different numbers (and types) of external lines. It is easy to
understand that this programme results in a set of graphs constructed
solely from minimal propagators and minimal effective vertices of
various loop orders
$l \leq L$
with different numbers and types of legs. Every such
($l$-th
order) vertex
$
V_{\ldots}^{l}(p_1,...,p_n)
$
with certain set of
$n$
legs takes the following typical form:
\begin{equation}
V_{\ldots}^{(l)}(p_1, \ldots , p_n) = \sum \limits_{a=1}^{M}
T_{\ldots}^{(a)} V_a^{(l)}({\nu}_1,...,{\nu}_{3n-10})\ ,
\label{8.1}
\end{equation}
where
$M$
is the number of relevant minimal tensor structures
$T_{\ldots}^{(a)}$
and
$V_a^{(l)}$
stands for the infinite formal series
\begin{equation}
V_a^{(l)}({\nu}_1, \ldots ,{\nu}_t) =
\sum\limits_{k_1,...,k_t = 0}^{\infty} V^{(a,l)}_{k_1 ... k_t}
{\nu}_1^{k_1} \ldots {\nu}_t^{k_t}\ ,\ \ \ \ \ \ \ \
t \equiv (3n-10)
\label{8.2}
\end{equation}
in powers of kinematical variables
(\ref{6.2}).

Clearly, the general form of
{\em minimal}
counterterm vertices of the loop order
$L$
under consideration looks precisely like that of
(\ref{8.1}), (\ref{8.2}).
The corresponding coefficients can be considered as the pieces of
those appearing in the expression
(\ref{8.2})
for the highest order
$l=L$
minimal effective vertex -- there is no necessity in writing them down
as special items. In turn, this means that, until we fix the set of
normalization prescriptions for minimal vertices, all the coefficients
$V^{(a,L)}_{k_1 ... k_t}$
in
(\ref{8.2})
should be taken as free parameters. We call them as
{\em the
$L$-th
level resultant parameters} (which are minimal by the very
construction). The only limitations for their values follow from the
requirement of finiteness of the
$L$-th
loop order amplitudes and the formal restrictions imposed by crossing
and Bose (Fermi) symmetry (until we fix the renormalization
prescriptions).

The important feature of the set of resultant parameters with
$l=0,1, \ldots ,L$
is that this set is
{\em full}
and
{\em closed}.
It is full because no other parameters are needed to compute all the
$S$-matrix
elements of the
$L$-th
order. It is closed in the sense that taking account of graphs with
$l > L$
loops leaves the lower level
($l \leq L$)
parameters  unchanged.

According to the results of
Sec.~\ref{sec-proof}, %
there is no need in formulating normalization prescriptions adjusting
finite parts of the coefficients at non-minimal counterterm vertices.
This means that, except the infinite parts needed to remove
divergences in subgraphs of the next loop order, those coefficients
can be chosen in a way most suitable for subsequent calculations. In
turn, this means that the full set of normalization conditions, needed
to fix the physical content of effective scattering theory, is not
larger than the set of corresponding resultant parameters.

Starting from this point we consider all the infinite renormalizations
done. Let us now briefly discuss the problems of convergence. In fact,
there are two problems closely connected with one another. The first
one is the problem of convergence of numerical series constructed from
the minimal parameters. Every coefficient in the form
(\ref{8.2})
for the resultant vertex presents an infinite sum of the parameters
describing individual secondary vertices. Since no one of those latter
parameters presents a measurable quantity, we do not think that the
problem of convergence of their infinite sums should be taken too
seriously.

Another problem is that of convergence of formal power series
presenting the resultant effective vertices. Let us first discuss the
case of tree-level resultant vertices with
$n = 4$
lines (recall that, irrespectively to a level, the resultant triple
vertices are just constants). Each one of the corresponding resultant
parameters
(see Fig.
\ref{4f})
\begin{figure}[ht]
\begin{center}
\begin{picture}(350,30)(25,-10)
\put(10,0){
\begin{picture}(100,20)(0,0) 
\put(0,-5){\shortstack{$\displaystyle\sum_{ \rm vertices \atop
\mbox{} }^{\infty}$}}
\put(40,0){\circle*{3}}
\put(40,0){\line(-1,1){10}}
\put(40,0){\line(-1,-1){10}}
\put(40,0){\line(1,-1){10}}
\put(40,0){\line(1,1){10}}
\put(60,-10){\shortstack{\boldmath$,$}}
\end{picture}
}

\put(100,0){
\begin{picture}(100,20)(0,0) 
\put(0,-5){\shortstack{$\displaystyle\sum_{ \rm vertices, \atop
\rm resonances }^{\infty}$}}
\put(40,0){\circle*{3}}
\put(40,0){\line(-1,1){10}}
\put(40,0){\line(-1,-1){10}}
\multiput(40,0)(1,0){20}{\circle*{2}}  
\put(45,-13){\shortstack{$R_s$}}
\put(60,0){\circle*{3}}
\put(60,0){\line(1,1){10}}
\put(60,0){\line(1,-1){10}}
\put(80,-10){\shortstack{\boldmath$,$}}
\end{picture}
}

\put(200,0){
\begin{picture}(100,20)(0,0) 
\put(0,-5){\shortstack{$\displaystyle\sum_{ \rm vertices, \atop
\rm resonances}^{\infty}$}}
\put(50,-10){\circle*{3}}
\put(50,-10){\line(-1,-1){10}}
\put(50,-10){\line(1,-1){10}}
\multiput(50,-10)(0,1){20}{\circle*{2}}  
\put(53,-5){\shortstack{$R_t$}}
\put(50,10){\circle*{3}}
\put(50,10){\line(-1,1){10}}
\put(50,10){\line(1,1){10}}
\put(75,-10){\shortstack{\boldmath$,$}}
\end{picture}
}

\put(300,0){
\begin{picture}(100,20)(0,0)  
\put(0,-5){\shortstack{$\displaystyle\sum_{ \rm vertices, \atop
\rm resonances}^{\infty}$}}
\put(50,-10){\circle*{3}}
\multiput(50,-10)(1,0){20}{\circle*{2}}  
\put(55,-23){\shortstack{$R_u$}}
\put(70,-10){\line(1,-1){10}}
\put(70,-10){\circle*{3}}
\put(40,-20){\line(1,1){15}}
\put(65,5){\line(1,1){10}}
\put(70,-10){\line(-1,1){25}}
\put(65,-5){\oval(20,20)[tl]}
\put(90,-10){\shortstack{\boldmath$.$}}
\end{picture} }

\end{picture}
\end{center}
\caption{Formal sum of graphs describing the tree-level amplitude
of the process
$2 \rightarrow 2$
before contracting the resonance lines.
$R_s$, $R_t$ and $R_u$ stand for all possible resonances in the
$s$-, $t$-, and $u$-channels, respectively.  The effective triple
vertices contain both minimal and non-minimal (with respect to inner
line) parameters.
\label{4f}}
\end{figure}
presents a sum of two items. The first item is just the relevant
minimal parameter appearing in the effective 4-vertex of the
Hamiltonian level. The second item stems from the reduction of graphs
with resonance exchanges in
$s$-, $t$- and $u$-channels.
It presents an infinite sum of products of the Hamiltonian triple
coupling constants, at least one of which being non-minimal with
respect to inner line. All the minimal triple couplings of the
Hamiltonian level are contained in the triple vertices describing the
pole parts of resonance contributions.

The resultant effective 4-vertex does not present an independent
element of Feynman rules: every time when it appears as a part of a
larger graph, one also has to take account of contributions due to the
resonance exchange graphs shown in
Fig.~\ref{4f}.
This note allows one to conclude that it makes no sense to discuss the
convergence of infinite series
(\ref{8.2})
for the resultant 4-vertex: only the full sum of tree-level graphs
under consideration must possess the desired convergency property.
This means that in the full sum of graphs (of a given loop order),
presenting an amplitude under consideration, we expect mutual
cancellations among various unwanted contributions which might occur
in every individual item.

Clearly, this argumentation equally applies to arbitrary effective
resultant vertex with
$n > 4$
lines as well as to the case of higher loop order vertices. Thus it
may happen that the resultant parameters describing the vertices with
different numbers of legs are not completely independent. Indeed, as
argued in
\cite{AVVVKS}
(the detailed analysis will be published elsewhere), the requirements
of convergence, crossing symmetry and polynomial boundedness lead to
highly nontrivial relations connecting the resultant parameters of the
vertices differing from one another by the number of legs.

\section{The essential parameters}
\label{sec-essential}
\mbox{}

In this Section we just give an idea on how to construct the essential
parameters from the resultant ones. The detailed analysis would
require too much space; it will be published elsewhere. A preliminary
discussion can be found in
\cite{AVVVKS}.

By way of illustration, let us consider the tree-level amplitude
describing a scattering process
$2 \rightarrow 2$.
For the following, it is convenient to consider in parallel three
different pairs of independent kinematical variables:
$$
[x, {\nu}_x],\ \ \ \ \ \ (x=s,t,u).
$$
Here
$s,t,u$
stand for the conventional Mandelstam variables, and
\begin{equation}
{\nu}_s \equiv (u-t),\ \ \ \ \ \
{\nu}_t \equiv (s-u),\ \ \ \ \ \
{\nu}_u \equiv (t-s).
\label{9.01}
\end{equation}
~From
(\ref{9.01})
it follows that
\begin{equation}
u = \frac{1}{2} ( 2\sigma - s + {\nu}_s ),\ \ \ \ \ \ %
t = \frac{1}{2} ( 2\sigma - s - {\nu}_s );
\label{9.02}
\end{equation}
\begin{equation}
s = \frac{1}{2} ( 2\sigma - t + {\nu}_t ),\ \ \ \ \ \ %
u = \frac{1}{2} ( 2\sigma - t - {\nu}_t );
\label{9.03}
\end{equation}
\begin{equation}
t = \frac{1}{2} ( 2\sigma - u + {\nu}_u ),\ \ \ \ \ \ %
s = \frac{1}{2} ( 2\sigma - u - {\nu}_u );
\label{9.04}
\end{equation}
where
$2 \sigma \equiv (m_1 + m_2 + m_3 + m_4)$
and
$m_i$ $(i=1,...,4)$
are the external particle masses.

The tree-level amplitude of the process under consideration is a sum
of four items each of which, in turn, presents an infinite sum of
contributions stemming either from the effective 4-vertex or from
graphs with resonance exchanges (see
Fig.~\ref{4f}).
In particular, the first term is an infinite sum of items each of
which originates from the corresponding Hamiltonian monomial
constructed from four field operators or/and their derivatives. It
takes a form of (formal!) infinite power series in two independent
kinematical variables. All the coefficients appearing in this series
are constructed from the corresponding minimal parameters of the
Hamiltonian level.

As to the triple vertices appearing in graphs with resonance
exchanges, they contain both minimal and non-minimal (with respect to
inner lines) parameters of the Hamiltonian level. The non-minimal
parameters do not contribute to the pole parts of graphs: as shown in
Sec.~\ref{sec-minimal},
they only contribute to smooth (`analytic') part. In contrast, all the
minimal parameters contribute to the values of residues at
corresponding poles. This means that, after the reduction of inner
lines, the amplitude can be presented in one of three equivalent forms
only differing from one another by the choice of variables:
\begin{equation}
M(s,{\nu}_s) =
\sum\limits_{i,j = 0}^{\infty} A^{(s)}_{ij} s^i {{\nu}_s}^j +
\sum\limits_{R_s}^{}
\frac{N_s^{(s)}({\nu}_s)}{s-M^2_R} +
\sum\limits_{R_t}^{}
\frac{N_t^{(s)}(s)}{{\nu}_s - ({\theta}_t - s)} +
\sum\limits_{R_u}^{}
\frac{N_u^{(s)}(s)}{{\nu}_s + ({\theta}_u - s)}\ ;
\label{9.6}
\end{equation}
\begin{equation}
M(t,{\nu}_t) =
\sum\limits_{i,j = 0}^{\infty} A^{(t)}_{ij} t^i {{\nu}_t}^j +
\sum\limits_{R_s}^{}
\frac{N_s^{(t)}(t)}{{\nu}_t + ( {\theta}_s - t)} +
\sum\limits_{R_t}^{}
\frac{N_t^{(t)}({\nu}_t)}{t - M^2_R} +
\sum\limits_{R_u}^{}
\frac{N_u^{(t)}(t)}{{\nu}_t - ({\theta}_u - t)}\ ;
\label{9.7}
\end{equation}
\begin{equation}
M(u,{\nu}_u) =
\sum\limits_{i,j = 0}^{\infty} A^{(u)}_{ij} u^i {{\nu}_u}^j +
\sum\limits_{R_s}^{}
\frac{N_s^{(u)}(u)}{{\nu}_u - ({\theta}_s - u)} +
\sum\limits_{R_t}^{}
\frac{N_t^{(u)}(u)}{{\nu}_u + ({\theta}_t - u)} +
\sum\limits_{R_u}^{}
\frac{N_u^{(u)}({\nu}_u)}{u - M^2_R}\ .
\label{9.8}
\end{equation}
Here
\begin{equation}
{\theta}_x \equiv (2 \sigma - M^2_{Rx})\ ,\ \ \ \ \
(x=s,t,u);
\label{9.9}
\end{equation}
the relations
(\ref{9.02}) -- (\ref{9.04})
have been used to rewrite denominators in terms of relevant pairs of
variables.

No one of the formal series
(\ref{9.6}) -- (\ref{9.8})
makes sense until we fix the order of summation and point out the
areas where we would like to assign meaning to those series. As argued
in
\cite{AVVV}
(see also
\cite{AVVVKS}),
it is natural to consider every series written in terms of the pair
$
[x, {\nu}_x]
$
in the corresponding thin 3-dimensional band (layer)
$$
B_x:\ \ \ \  \left\{ x \in {\bf R},\  {\nu}_x \in {\bf C};\
           x \in (-\epsilon, \epsilon) \right\}\ \ \ \ \ \
(x=s,t,u).
$$
By condition, the thickness
$2\epsilon$
of the layer
$B_x$
should be taken sufficiently small such that
$\epsilon < min \{ M_{Rx}^2 \}$.
This means that those items which contain fixed (independent of
${\nu}_x$)
poles in
$x$
do not result in singular contributions in
$B_x$.
Hence, in
$B_x$
the expression for the amplitude can be rewritten as formal sum of
contributions due to sliding (depending on
$x$)
poles in
${\nu}_x$
plus the term which is formally regular in both variables. For
example, in
$B_u$
we have:
\begin{equation}
M(u,{\nu}_u) =
\sum\limits_{i,j = 0}^{\infty} M^{(u)}_{ij} u^i {{\nu}_u}^j +
\sum\limits_{R_s}^{}
\frac{N_s^{(u)}(u)}{{\nu}_u - ({\theta}_s - u)} +
\sum\limits_{R_t}^{}
\frac{N_t^{(u)}(u)}{{\nu}_u + ({\theta}_t - u)}\ ,\ \ \ \ \
([u, {\nu}_u] \in B_u).
\label{9.10}
\end{equation}
The corresponding formal expressions for the amplitude in
$B_s$
and
$B_t$
can be rewritten precisely in the same way. We would like to stress
that every coefficient in
(\ref{9.10})
is constructed from the tree level resultant parameters.

The special convenience of the form
(\ref{9.10})
is explained by the following reason. At every fixed
$u \in B_u$
this form can be treated as a uniformly converging series presenting a
meromorphic function of one complex variable
${\nu}_u$
and one real parameter
$u$.
The possibility of such interpretation is provided by the general
theorem (due to Mittag-Leffler) known from complex analysis (see,
e.g.,
\cite{Shabat}, \cite{Arfken}).
To make use of this theorem in its constructive form, one has to
impose certain limitations on the values of resultant parameters.
Besides, in order to provide a guarantee that the amplitude possesses
desired properties of crossing symmetry, one needs to consider in
parallel three different forms of the type
(\ref{9.10})
(in
$B_s$, $B_t$
and
$B_u$).
In the domains of mutual intersections
$$
D_s \equiv B_t \cap B_u,\ \ \ \ \
D_t \equiv B_u \cap B_s,\ \ \ \ \
D_u \equiv B_s \cap B_t,
$$
the corresponding forms must identically coincide in pairs. This
requirement leads to additional limitations%
\footnote{Called in
\cite{VV} -- \cite{AVVVKS}
as bootstrap equations.} %
strongly restricting the allowed values of resultant parameters.
Those limitations take a form of an infinite system of algebraic
equations connecting different parameters among themselves and, hence,
reducing the number of
{\em independent}
parameters needed to fix a particular effective scattering theory. The
full set of independent combinations of resultant parameters can be
considered as the set of true essential parameters which require
formulating the renormalization prescriptions.

\section{Conclusion}
\label{sec-conclusion}
\mbox{}

The main result of the above analysis can be formulated as follows.
{\em To describe the scattering processes in the framework of an
effective field theory one has no need in fixing the detailed
structure of particle interactions off the mass shell. All the
information needed to fix the numerical values of
$S$-matrix
elements at a given loop order
$L$
is contained in the values of the resultant parameters of
$L$-th
and lower levels.}
This result coincides with that obtained by S.~Weinberg, M.~Scadron
and J.~Wright in series of papers
\cite{Quasi1} -- \cite{Quasi5}
on nonrelativistic scattering theory.

The central idea of our work is that the number of
{\em independent}
normalization prescriptions needed to fix the physical content of an
effective scattering theory is much less than the total number of
resultant parameters. As explained in
Sec.~\ref{sec-essential},
certain natural consistency requirements lead to an infinite number of
constraints strongly restricting the allowed physical values of those
parameters. This point will be discussed in detail in the next
article.

\section*{Acknowledgements}

We are grateful to K.~Semenov-Tian-Shanski, A.~Vasiliev, M.~Vyazovski,
V.~Cheianov, H.~Nielsen and J.~Schechter for stimulating discussions.

The work was supported in part by INTAS (project 587, 2000),
RFBR (grant 01-02-17152) and by Ministry of Education of Russia
(grant E00-3.3-208). The work by A.~Vereshagin was supported by
Meltzers H\o yskolefond (Studentprosjektstipend 2002).

\end{document}